\documentclass[aps,epsf,twocolumn]{revtex4}
\usepackage{graphicx}
\usepackage{setspace}
\begin{document}
\title{Equation of state in a small system: Violation of an assumption of Maxwell's demon}
\author{T. Hondou\footnote{E-mail: hondou@cmpt.phys.tohoku.ac.jp}}
\address{Department of Physics, Tohoku University, Sendai 980-8578, Japan \\
            Institut Curie, Section Recherche, UMR168, 26 rue d'ULM, 75248 Paris Cedex 05, France}
\begin{abstract}

An equation of state for an ideal gas with a small number of particles is studied. 
The resulting equation is found to differ from that expected in conventional thermodynamics,
which is strikingly illustrated when considering the traditional thermodynamic problem of Maxwell's demon. We clarify the mechanism of this different feature of thermodynamics arising in small systems.

\end{abstract}
\maketitle

 The thought experiment by Maxwell\cite{Maxwell} known as Maxwell's demon
 has attracted the attention of
 physicists for more than a century\cite{review}. Maxwell's demon is a well-known paradox
based on the fundamentals of thermodynamics, which seemingly permits a violation of the 
second law of thermodynamics. 
In the original thought experiment, we are asked to imagine a demon that controls a gate between two compartments. Being able to determine the speed of the particles in each compartment, the demon opens the gate in such a way as to collect more high-speed particles in one compartment, thus decreasing the entropy of the system, in seeming violation of the second law.
 About half a century after Maxwell, 
 Szilard extended Maxwell's demon, devising a model which allowed quantitative description on the extracted work\cite{Szilard}. 

 The framework of Szilard's formulation is depicted in Fig.1. One thermal molecule is in
an isothermal system and the cycle of Maxwell's demon
is considered to be composed of  two processes:  1) measurement (observation) of the
molecule position and 2) isothermal expansion in the process from State B to C.
 By measurement, the {\em demon} determines whether the molecule is on the left or right side
and inserts a wall to create a {\em piston},
which is accompanied by a mechanical load (weight) for extraction of work. In this way, work of an amount $W = k_{\rm B}T \ln 2$ can be extracted from the system by isothermal expansion.
Through cyclic operation, i.e. following the process from States
A $\rightarrow$ B $\rightarrow$ C $\rightarrow$ A,  
work is perpetually extracted out of the isothermal system. 
The operation transfers ambient thermal energy into work through
the isothermal expansion of the ``piston", resulting in a decrease in entropy, $k_{\rm B} \ln 2$. 
In order to save the second law,
Szilard described how the entropy decrease would be compensated
``... if  the execution of such a measurement were, for instance, always accompanied by
production of $k \ln 2$ units of entropy"\cite{Szilard}.

\begin{figure}
\includegraphics[scale=0.4]{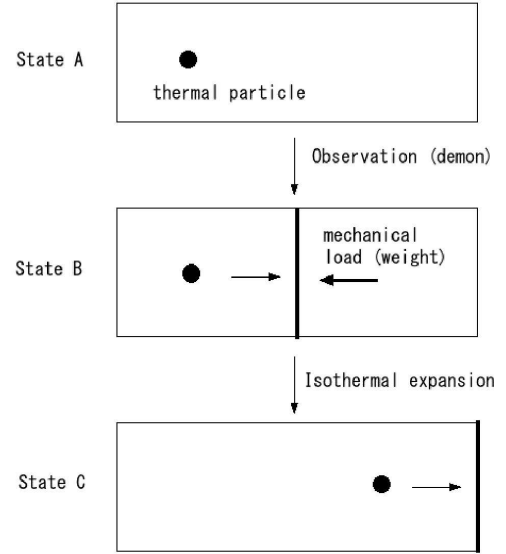}
\caption{Conventional scheme of Maxwell's demon, after Szilard (1929)}
\end{figure}

As has recently been pointed out\cite{Magnasco}, Szilard did not prove his compensation theory, but assumed that his heat engine would always preserve the second law.
Most studies of Maxwell's demon have followed the theoretical framework of Szilard\cite{review}, and have focused on the energy cost of observation
\cite{11}.  In such studies,
it is generally assumed that, for ideal gases, one can extract work, $W$, even for a one-molecule isothermal piston, of an amount given by the relation
$W=\int_{V_i}^{V_f} P(V) dV = n R T \ln{(V_f/V_i)}$, where $P$ is the pressure, $n$ is the number of molecules in a mole, $R$ is the gas constant, $T$ is the temperature, and $V_{i}$, and $V_f$ are the initial and final volumes of the piston,
respectively\cite{22}. Thus, the conventional equation of state 
for an ideal gas has been implicitly assumed:
$PV = n RT$.
In other words, studies of Maxwell's demon have generally been performed within the framework
 of conventional 
{\it macroscopic} thermodynamics, regardless of the small number of molecules\cite{Hill}.
 
On the other hand, recent experimental developments have enabled experimental 
observation of even a single molecule
in some situations. In molecular motors\cite{Finer,Miyata,Molloy} the value of the work extracted 
in an elementary process fluctuates over repeated experiments, which is in contrast to the thought experiment of Maxwell's demon.
This suggests that there may be an essential fault in the basic assumption of Maxwell's demon on this scale, because an essential experimentally observed feature of a small-scale system, thermal fluctuation, has not been properly 
accounted for.
In addition, recent studies that describe
the energetics in thermally fluctuating systems have shown
that there is another way of describing the thermodynamics of 
small systems kinetically without employing
entropy\cite{KenJPS}. 
 From this background, a question was recently raised by Hatano and Sasa\cite{Hatano} as to the validity of the assumption of the work extracted
in Szilard's model.  The study was an attempt at 
quantitative estimation of the work extracted
in an operational manner, which is in contrast to conventional thought experiments.  
Although their study was a pioneering one, the model resulted 
in broken detailed balance even in the equilibrium state.

In this paper, we analyze the equation of state in an extremely small system,
 within the boundary of kinetics. We discuss an equilibrium process here; however, an interesting aspect of relaxation into equilibrium is discussed by Crosignani and Porto\cite{Crosignani}.
We show that the equation of state that Szilard and others have assumed for isothermal expansion
is not appropriate. 
We clarify the theoretical origin of the difference between an extremely small system and conventional macroscopic systems
towards the construction of a system of thermodynamics for small 
systems\cite{Hill}. 

\begin{figure}
\includegraphics[scale=0.6]{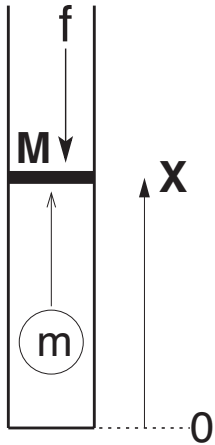}
\caption{One-dimensional system composed of thermal particles and a movable piston. 
The force of a constant load, $f$, on the movable piston competes with the
 ``pressure" of thermal particles.}
\end{figure}

 We begin with an estimation of the equation of state of {\it one} molecule because
 typical features of small systems are expected at this limit.
In order to derive the equation, we must confine ourselves to a concrete model that permits us to treat 
the system only
within the framework of kinetics.
As in the study by Hatano and Sasa\cite{Hatano},  we consider a one-dimensional
piston in thermal equilibrium (Fig. 2), where the dynamical variables are
 $N$ ``thermal particles" and a movable piston on one side.
By replacing the volume $V$ 
by the distance between the wall and the movable piston $X$, 
 $n$ by  $N/n_{\rm a}$, and $R$ by $n_{\rm a} k_{\rm B}$, we 
obtain the equation of state for the system, 
$P X = N k_{\rm B} T$, where $n_{\rm a}$ and $k_{\rm B}$ are the Avogadro constant and the Boltzmann constant, respectively.

Hatano and Sasa claimed that for ideal gases the 
equation of state depends on the masses of the thermal particles and the 
piston and that the conventional equation of state is valid if the 
ratio of the
mass of the piston to that of 
the particle is sufficiently large. However, we found, by the method of stochastic energetics\cite{KenJPS},
that in the study by Hatano
and Sasa, there is a finite current between two thermal baths, one of which 
is in contact with a thermal wall and the other of which is in contact with a 
movable piston, resulting in the broken detailed balance. We can learn from this that 
we cannot naively utilize a ``hybrid" model, for example,   
Langevin dynamics (for a piston) and Hamiltonian dynamics (for a thermal particle)  
simultaneously,
because Langevin dynamics is 
a coarse-grained description of Hamiltonian dynamics\cite{Mori}.
In order to maintain self-consistency in a class of description,
 we should return to the framework of elementary mechanics.

In most literature on Maxwell's demon, it is assumed
that we {\it can fix} a movable piston 
by applying an external force of the same 
value as the conventional (internal) {\it pressure}. This assumption fails in a small system, as discussed later. Thus, 
we replace the coarse-grained quantity ``pressure" $P$ by the mechanical force 
$f$ externally exerted on the 
piston. This yields a reference equation,
  $f X = N k_{\rm B} T$, 
which we will use to discuss the validity of the conventional equation of state,
and hence the validity of the frameworks of the previous studies. We assume that collisions between the variables
 (particle(s) and a  piston) are 
 perfectly elastic and that the movable piston is frictionless,
 since ``friction" is also a coarse-grained quantity
 which must be excluded for consistency. 
  Then, the total energy of the system of one thermal particle ($N=1$) is written as

\begin{equation}
E=\frac{p_{\rm t}^{2}}{2m} + \frac{p_{\rm p}^{2}}{2M} + f X ,
\end{equation}
where $m (p_{\rm t}) $ and  $M (p_{\rm p})$ are the masses (momenta) of the thermal particle and the piston, respectively, and
$f$ is a force applied to the movable piston. 
To keep the system isothermal, a thermal wall\cite{Bergmann} is introduced.
The particle
is reflected with a positive random velocity
according to Maxwell's velocity distribution at the boundary of the system,
$x=0$, where $x$ is the position of a thermal particle.
The thermal wall realizes thermal equilibrium
of the system. 
Because the system is 
one-dimensional, the thermal particle
is always lower than the piston, $x \leq X$. 

One naively expects that the equation of state for 
one molecule should be obtained by replacing $N$ by
$1$ in the reference equation $f X = N k_{\rm B} T$; 
namely $f X = k_{\rm B} T$. The extension of conventional thermodynamics into those of one molecule
 has been performed since 
Szilard\cite{Szilard}.
In the following, we will investigate the validity of this extension in order to uncover the nature of thermodynamics for a few molecules. 
Three methods of analyses are performed independently: 1) numerical simulation, 2) Master equation analysis, and 3) Gibbs' statistical mechanics.

In the simulation, the system obeys Hamiltonian dynamics with the exception of the thermal wall, in which 
 the thermal particle is reflected with a random velocity within the velocity distribution
 $ \frac{m |v|}{k_B T} e^{-m v^2 / 2 k_B T} \,  \, 
(v>0)$ when the particle arrives at the boundary $x=0$.
The condition of the thermal wall is known to direct the target system into thermal equilibrium\cite{Bergmann}.
The numerical simulation revealed that the equation of state for one molecule is in fact 
\begin{equation}
f \langle X \rangle = 2 k_{\rm B} T ,
\label{eq-double}
\end{equation}
where the angular brackets indicate an average value.
This result is obviously different from that which is conventionally
assumed.

\begin{spacing}{1}
We obtained the same analytical result using the Master equation of the distribution
function in a phase space, which consists of Liouville terms and collision terms.
Let $\rho(x, v, X, v_p)$  be the probability density of the particles (a thermal particle
and a movable piston), and $v$ and $v_p$ the velocities  of the thermal particle and movable
piston, respectively.
Here we use velocities instead of 
momenta for simplicity.  The density function obeys the following equation:
\end{spacing}
{\small
\begin{eqnarray}
\lefteqn{\frac{\partial \rho(x, v, X, v_p)}{\partial t} =}  \nonumber \\
& & - v \frac{\partial \rho}{\partial x} - v_p \frac{ \partial\rho}{\partial X}
  +\frac{f}{M} \frac{\partial \rho}{\partial v_p} \nonumber
\\
  &    & - \theta (v-v_p) (v-v_p) \rho(x, v, X, v_p) \delta (x -
X) \nonumber
\\
&   & + \theta (v_p-v) (v_p-v)   \nonumber
\\
& & \times  \rho(x, \frac{m-M}{M+m} v
+\frac{2M}{M+m} v_p, X, \frac{2m}{M+m} v + \frac{M-m}{M+m} v_p)  \nonumber
\\
& & \delta(x - X) .
\label{rho11}
\end{eqnarray}
}
The first three terms of the R.H.S. of this equation come from the Liouville equation and  the last two terms come from collision effects.
A stochastic boundary condition is applied to the system at one end, that is, $\rho(x=+0, v) =\frac{f}{T} \frac{\sqrt{m}}{\sqrt{2 \pi T}}\exp[-mv^2/2T]$ for
 $v > 0$. The stochastic condition corresponds to the thermal wall in the numerical simulation.
 From straightforward calculation, one obtains a stationary solution,
$\rho (x, p_{\rm t}, X, p_{\rm p}) = \frac{f^2}{2 \pi (k_{\rm B} T)^3 \sqrt{mM}} \exp\{-
(\frac{p_{\rm t}^{2}/2m + p_{\rm p}^2 /2M + f X  }{k_{\rm B} T}) \}   \theta
(X-x) ,$
where $\theta$ is a Heaviside step function.
From this equation, we again obtain Eq. (\ref{eq-double}).
Note that the result is independent of the masses of both the thermal particle 
and the movable piston, which is in contrast to the result by Hatano-Sasa \cite{Hatano}.

\begin{figure}
\includegraphics[scale=0.6]{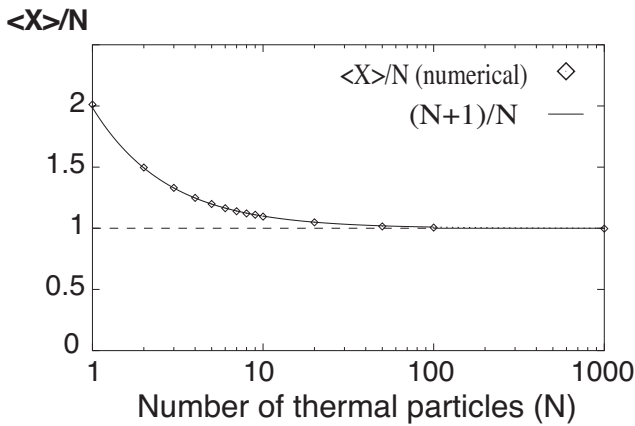}
\caption{The dependence of the number of particles, $N$, on the time average of the position of the piston,
where $k_{\rm B} = T = f = 1$ for simplicity.
For comparison, the average position is normalized by $N$.
If the reference equation of state, $f X = N k_{\rm B} T$, {\it were} valid for {\it any} $N$, $\langle X \rangle / N$ would
remain as 1.
}
\label{fig:N}
\end{figure}

Although we introduced a thermal wall and, correspondingly, a stochastic
boundary condition, one may obtain the same generalized result
without applying such boundary conditions, but instead using conventional Gibbs'
statistical mechanics. Since the present system has two constant parameters,
 temperature and force (applied to the piston),
the system can be described by a pressure ensemble (P-T ensemble) of 
Gibb's statistical mechanics\cite{Kubo}. 
The quantity ``pressure" in a conventional {\it pressure} ensemble is replaced here by a force. This replacement does 
not alter any general result of the pressure ensemble,
because one can reprove any derivation in the pressure ensemble with
this replacement. The Laplace transformation between a partition function of the canonical ensemble $Z$
and that of the pressure ensemble $Y$ is given as
$ Y= \int_{0}^{\infty} dV \exp{[-PV/(k_{\rm B} T)]} Z  $\cite{Kubo}, from which
we obtain the partition function of the pressure ensemble of the system
of $N$ particles, 
$Y  = F (\frac{k_{\rm B} T}{f})^{N+1} $,
where $F$ is a factor that will drop out later. 
The average position of the movable piston is obtained as
$\langle X \rangle = - \frac{1}{k_{\rm B}T} \frac{\partial \ln Y}{\partial f} = (N+1) k_{\rm B} T/ f .$
Thus, we have 
$f \langle X \rangle  = (N+1) k_{\rm B} T.$
The previous result, $f \langle X \rangle  = 2 k_{\rm B} T$, is again obtained by replacing $N$ by 1.
One finds that in a sufficiently large system, $N \gg 1$ (thermodynamic limit),
 $N+1$ may be replaced by $N$.
However, 
in the small system, the difference between $N+1$ and $N$ is crucial, as shown in Fig. \ref{fig:N}.

This result 
is apparently in contrast to the basic assumption made by Szilard and others,
in which the macroscopic thermodynamic relation is assumed to hold even
in a small system, namely, $f X = k_{\rm B} T$ for one molecule.
In such studies, we find that it has been assumed that one can ``fix" or quasi-statistically ``control"
the position of the piston, where the existence of the working bath (or "working reservoir")\cite{55} behind the piston
 is {\it implicitly} assumed. 
However, for a small system, the question as to how one can fix or control the piston in an operational
or experimental manner should be carefully examined\cite{KenPRE}.
The validity of the concept of the working bath
is not evident in small systems\cite{KenJPS}.
If one fixes the position of the piston, that is, if one eliminates the variability
of the piston, work cannot be extracted out of the system 
because the ``fixed" piston cannot transfer thermal energy into work. Such a framework is in contradiction
to, and thus is no longer physically valid in, studies of Maxwell's demon. 
This is precisely the reason why we introduced an 
external load to the piston instead of using the concepts of ``pressure" or a ``working bath".
It should also be noted that the movable piston satisfies the thermal property (Maxwell's velocity
distribution) merely through kinetic collision with a thermal particle.
The evidence supports
the physical consistency of the present description.

\begin{figure}
\includegraphics[scale=0.55]{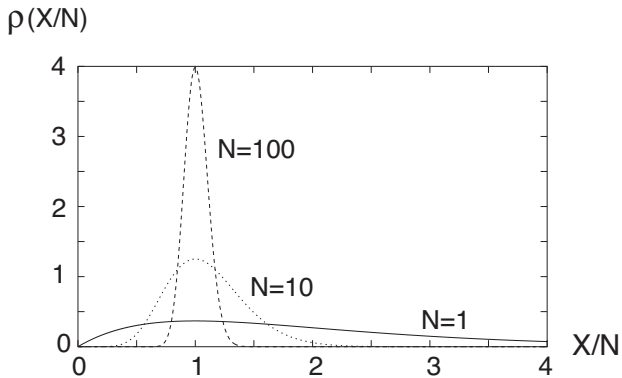}
\caption{Distribution function of the position of the movable piston,
where $k_{\rm B} = T = f = 1$ and the horizontal axis is normalized by $N$ for simplicity.
As the number of particles decreases, the deviation from the most probable position,
$ X^{\ast}  / N = 1$, becomes substantial.
For comparison, the position is normalized by the number of thermal 
particles,  $N$.}
\label{fig.p}
\end{figure}

In the following paragraphs, we will present the role of thermal fluctuation as it essentially 
relates to the equation of state\cite{0807}.
We have the distribution function in phase space,
$\rho =  F' e^{-f X/k_{\rm B} T }
\theta (X - x_{N}) \theta(x_{N} - x_{N-1}) \cdots \theta(x_2 - x_1),$
where $F'$ and $x_i$ are a factor and
the position of the $i$-th thermal particle ($X \geq x_N 
\geq \cdots \geq x_2 \geq x_1$), respectively.
By integrating all the spatial coordinates, except for that of the movable piston, $X$,
we obtain the distribution function of the movable piston: 
{\small
\begin{equation}
 \rho(X) 
 =  \frac{1}{N!} (\frac{k_{\rm B}T}{f})^{N+1}  \exp{\{-\frac{f X - N k_{\rm B} T \ln{X}}{k_{\rm B} T}\}}.
\label{rho}
\end{equation}
}
Now, we define the coordinate $X=X^{*}$ where the distribution function
 $\rho(X) $ is 
maximum. From Eq. (\ref{rho}), we get the relation
 $ X^{*} = N k_{\rm B} T/f $.
Thus, the conventional thermodynamic relation $f X = N k_{\rm B} T$ 
still remains for the most probable position of the piston, $X^{\ast}$,  although it does  not hold for the average position. 
The reason why the average position 
of the piston deviates  from the most probable position, $X^{\ast}$, 
is that thermal fluctuation of the piston is substantial for small systems,
as seen by the relation
$\sqrt{ \langle (X - \langle X \rangle )^2 \rangle }/\langle X  \rangle = 1/\sqrt{N+1}  .$
Figure 4 shows the distribution function of the movable piston for three different
 quantities
of thermal particles ($N=1$, $10$, and $100$), where the horizontal axis is 
normalized by $N$. As the number of particles decreases, the fluctuation from the most probable
position, $X=X^{\ast}$, and simultaneously
the asymmetry of the distribution function, increase, while the most probable position of the piston
continues to obey the conventional equation of state. Thus, the {\em average} equation of state 
deviates from the conventional one as the number of particles decreases.
An interesting question relevant to any technique that reduces the thermal fluctuation of one molecule, as applied in probe microscopy\cite{Yanagida}, may arise here: Which position does the frozen position of a molecule correspond to, the most probable position or the average position obtained before the reduction in thermal fluctuation?

We can learn here the features of the concept of a ``
quasi-static process" in extremely small  systems.
Szilard and others implicitly assumed that the expansion process could be described
in the same quasi-static framework as its macroscopic counterpart.
However, the condition of a quasi-static process requires a
 completely different framework at the microscopic level from that
 at the macroscopic level\cite{KenPRE}. 
A quasi-static condition always 
requires that a system should itinerate all the possible
phase space sufficiently under the current parameters of the system.
For this requirement, the system parameters should vary slowly
enough during a quasi-static thermodynamic process.
In our case, what should vary slowly is {\em not} the position of the piston but the force, $f$.
In the expansion process, the
piston moves rapidly following a slow change in the force. Thus, the
conventional picture of a gradually expanding piston does not
correspond to quasi-static expansion in small systems.
Szilard and others had not anticipated this fact.
  
In this paper, we showed in the framework of Szilard's model that
the equation of state   for a few molecules is 
different from that previously assumed by Szilard and others.
 In the thermodynamic limit, the equation of state is found to coincide with the conventional equation of state\cite{0809}.
The result reveals that the macroscopic framework of thermodynamics has been incorrectly extended into the one molecule
world, which has been assumed in the literature of Maxwell's demon.

 This counterexample to the basic assumption of Maxwell's demon brings 
 to light the need
 to develop the thermodynamics of small systems\cite{KenJPS}. This is important not only for Maxwell's demon but also for proper interpretation and modeling of 
one-molecule experiments,
in which the experimental data vary among observations. One-molecule
 thermodynamics is {\it not} a rescaled version of its macroscopic
  counterpart\cite{Evans}. Filling the gap between one-molecule 
  thermodynamics 
and macroscopic thermodynamics should be an interesting challenge from 
both philosophical 
and practical viewpoints of physics.

 The author would like to thank K. Sekimoto for making him aware of the problem
 and introducing him to the method of Master equations, J. Lebowitz for useful comments on the pressure ensemble
 and other fruitful discussions, and S. Sasa, M. Kikuchi, M. Sano, Y. Hayakawa, S. Takagi, 
F. J\"ulicher, J. Prost, T. Kawakatsu for helpful comments. This work was supported in part by
the Inamori Foundation, and by a Grand-in-Aid from JSPS (Nos. 60261575 and 17654082).


\begin{thebibliography}{99}
\baselineskip 10mm
\bibitem{Maxwell}
Maxwell J. C., {\em Theory of Heat} (Longmans, London) 1871.

\bibitem{review}
Leff H. S. and Rex A. F.  (Eds.), {\em Maxwell's Demon} (Adam Hilger, Bristol) 1990.

\bibitem{Szilard}
Szilard L., Z. Phys. {\bf 53} (1929) 840; An English translation is found in ref. \cite{review}.

\bibitem{Magnasco}
Magnasco M. O. , Europhys. Lett. {\bf 33} (1996) 583.

\bibitem{11}
Brillouin L., J. Appl. Phys. {\bf 22} (1951) 334;
Rothstein J., Science {\bf 114} (1951) 171.

\bibitem{22}
For example, 
Raymond R. C. , Am. J. Phys. {\bf 19}, (1951) 109;
Finfgeld C. and Machlup S., Am. J. Phys. {\bf 28} (1960)  324.

\bibitem{Hill}
Hill T. L. , {\em Thermodynamics of Small Systems} (W.A. Benjamin, New York) 1963. 

\bibitem{Finer} Finer J. T., Simmons R. M., and Spudich J. A., Nature {\bf 368} (1994) 113.


\bibitem{Miyata}
 H. Miyata et al.,
J. Biochem. {\bf 115} (1994) 644.

\bibitem{Molloy}
 J. E. Molloy et al., Nature {\bf 378} (1995) 209.

\bibitem{KenJPS}
Sekimoto K., J. Phys. Soc. Jpn. {\bf 66} (1997) 1234.


\bibitem{0807}
Derivations of Landauer's principle, including thermal fluctuation, are found in,  Piechocinska, B., Phys. Rev. A {\bf 61} (2000) 062314;  Kawai, R., Parrondo, J. M. R. and Broeck, C. V., Phys. Rev. Lett. {\bf 98} (2007), 080602.

\bibitem{Hatano}
Hatano T. and Sasa S., Prog. Theor. Phys. {\bf 100} (1998) 695. 

\bibitem{Crosignani}
Crosignani B. and Porto P. D., Europhys. Lett. {\bf 53} (2001) 290.

\bibitem{Mori}
Mori H., Prog. Theor. Phys. {\bf 33} (1965) 423.

\bibitem{Bergmann}
Salwen H., Sadowski W., and Bergmann P. G., Bull. Am. Phys. Soc. Ser. II {\bf 1} (1956) 221.

\bibitem{Kubo}
Toda M. et al., {\em Statistical Physics I:
Equilibrium Statistical Mechanics} (2nd ed.) (Springer, Berlin) 1991.

\bibitem{55}
Callen H. B., {\em Thermodynamics} (John Wiley, New York) 1960.

\bibitem{KenPRE}
Sekimoto K., Takagi F. and Hondou T., Phys. Rev. E {\bf 62} (2002) 7759.

\bibitem{Yanagida}
Aoki T. et al., Ultramicroscopy {\bf 70} (1997) 45.

\bibitem{0809}
The extracted work per molecule is also the same as the conventional one, $k_{\rm B} T  \ln 2$, in the thermodynamic limit. 

\bibitem{Evans}
Wang G. M. et al., Phys. Rev. Lett. {\bf 89} (2002) 050601.

\end{thebibliography}
\end{document}